\documentclass[apjl]{emulateapj}

\usepackage{graphicx}
\usepackage{setspace}

\slugcomment{Accepted to ApJ Letters January 3rd 2017}

\shorttitle{A Cloudy Atmosphere for WASP-101b}
\shortauthors{Wakeford et al.}

\begin{document}


\title{HST P\lowercase{an}CET program: A Cloudy Atmosphere for the promising JWST target WASP-101\lowercase{b}}



\author{H.R. Wakeford\altaffilmark{1}, 
K.B. Stevenson\altaffilmark{2}, 
N.K. Lewis\altaffilmark{2}, 
D.K. Sing\altaffilmark{3}, 
M. L\'{o}pez-Morales\altaffilmark{4}, 
M. Marley\altaffilmark{5}, 
T. Kataria\altaffilmark{6}, 
A. Mandell\altaffilmark{1}, 
    G.E. Ballester\altaffilmark{7},
    J. Barstow\altaffilmark{8},
    L. Ben-Jaffel\altaffilmark{9},
    V. Bourrier\altaffilmark{10},
    L.A. Buchhave\altaffilmark{11},
    D. Ehrenreich\altaffilmark{10},
    T.M. Evans\altaffilmark{3},
    A. Garc\'ia Mu\~noz\altaffilmark{12},
    G. Henry\altaffilmark{13},
    H. Knutson\altaffilmark{14},
    P. Lavvas\altaffilmark{15},
    A. Lecavelier des Etangs\altaffilmark{9},
    N. Nikolov\altaffilmark{3},
    J. Sanz-Forcada\altaffilmark{16}.
}

\affil{1. Planetary Systems Lab, NASA Goddard Space Flight Center, Greenbelt, MD 20771, USA}
\email{hannah.wakeford@nasa.gov}
\affil{2. Space Telescope Science Institute, 3700 San Martin Drive, Baltimore, MD 21218, USA}
\affil{3. Astrophysics Group, Physics Building, University of Exeter, Stocker Road, Exeter EX4 4QL, UK}
\affil{4. Harvard-Smithsonian Center for Astrophysics, Cambridge, MA 02138, USA}
\affil{5. NASA Ames Research Center, MS 245-5, Moffett Field, CA 94035, USA}
\affil{6. NASA Jet Propulsion Laboratory, 4800 Oak Grove Dr, Pasadena, CA 91109, USA}
\affil{7. Dept. of Planetary Sciences \& Lunar \& Planetary Lab., University of Arizona, 1541 E Univ. Blvd., Tucson, AZ 85721, USA}
\affil{8. Physics and Astronomy, University College London, London, UK}
\affil{9. Institut d’Astrophysique de Paris, CNRS, UMR~7095 \& Sorbonne Universit\'es, UPMC Paris 6, 98 bis bd Arago, 75014 Paris, France}
\affil{10. Observatoire de l'Universit\'e de Gen\`eve, 51 chemin des Maillettes, 1290 Sauverny, Switzerland}
\affil{11. Centre for Star and Planet Formation, Niels Bohr Institute and Natural History Museum \&, University of Copenhagen, \O{}ster Voldgade 5-7, DK-1350 Copenhagen K, Denmark }
\affil{12. Zentrum f\"ur Astronomie und Astrophysik, Technische Universit\"at Berlin, D-10623 Berlin, Germany}
\affil{13. Center of Excellence in Information Systems, Tennessee State University, Nashville, TN  37209, USA}
\affil{14. Division of Geological and Planetary Sciences, California Institute of Technology, Pasadena, CA 91125, USA}
\affil{15. Groupe de Spectroscopie Mol\'eculaire et Atmosph\'erique, Universit\'e de Reims, Champagne-Ardenne, CNRS UMR 7331, France}
\affil{16. Centro de Astrobiolog\'{i}a (CSIC-INTA), ESAC Campus, P.O. Box 78, E-28691 Villanueva de la Ca\~{n}ada, Madrid, Spain}




\begin{abstract}
We present results from the first observations of the Hubble Space Telescope (HST) Panchromatic Comparative Exoplanet Treasury (PanCET) program for WASP-101b, a highly inflated hot Jupiter and one of the community targets proposed for the James Webb Space Telescope (JWST) Early Release Science (ERS) program. From a single HST Wide Field Camera 3 (WFC3) observation, we find that the near-infrared transmission spectrum of WASP-101b contains no significant H$_2$O absorption features and we rule out a clear atmosphere at 13$\sigma$. Therefore, WASP-101b is not an optimum target for a JWST ERS program aimed at observing strong molecular transmission features. We compare WASP-101b to the well studied and nearly identical hot Jupiter WASP-31b. These twin planets show similar temperature-pressure profiles and atmospheric features in the near-infrared. We suggest exoplanets in the same parameter space as WASP-101b and WASP-31b will also exhibit cloudy transmission spectral features. For future HST exoplanet studies, our analysis also suggests that a lower count limit needs to be exceeded per pixel on the detector in order to avoid unwanted instrumental systematics.
\end{abstract}




\section{Introduction} 
Observations of hot Jupiters with Hubble Space Telescope (HST) have revealed a diversity of atmospheres from clear to cloudy (\citealt{Sing2016}). In the near-infrared with Wide Field Camera 3 (WFC3), significant water absorption features have been measured for a number of exoplanets (e.g., \citealt{deming2013,wakeford2013,stevenson2014a,evans2016}). However, a significant number also show little or no evidence of water absorption (e.g., \citealt{line2013,ehrenreich2014,knutson2014a,Sing2015a}), which is plausibly due to clouds high in the planetary atmosphere (\citealt{Sing2016}). A number of studies across temperature, mass, and metallicity ranges for exoplanets have shown that clouds and aerosols will likely play a critical role in exoplanetary atmospheres (e.g., \citealt{morley2015,wakeford2015,parmentier2016,wakeford2016b}). Several studies, such as by \citet{greene2016}, have shown that with a cloud-free atmosphere key constraints can be placed on the abundance of both water and carbon-species, important in breaking the C/O-metallicity dependence. This in turn can give insights into planetary formation (e.g., \citealt{fortney2013,kreidberg2014b,benneke2015}). Therefore, understanding the presence or absence of clouds in the atmosphere of exoplanets is important in the lead up to the launch of the James Webb Space Telescope (JWST). 

\begin{figure}
\centering 
  \includegraphics[width=0.45\textwidth]{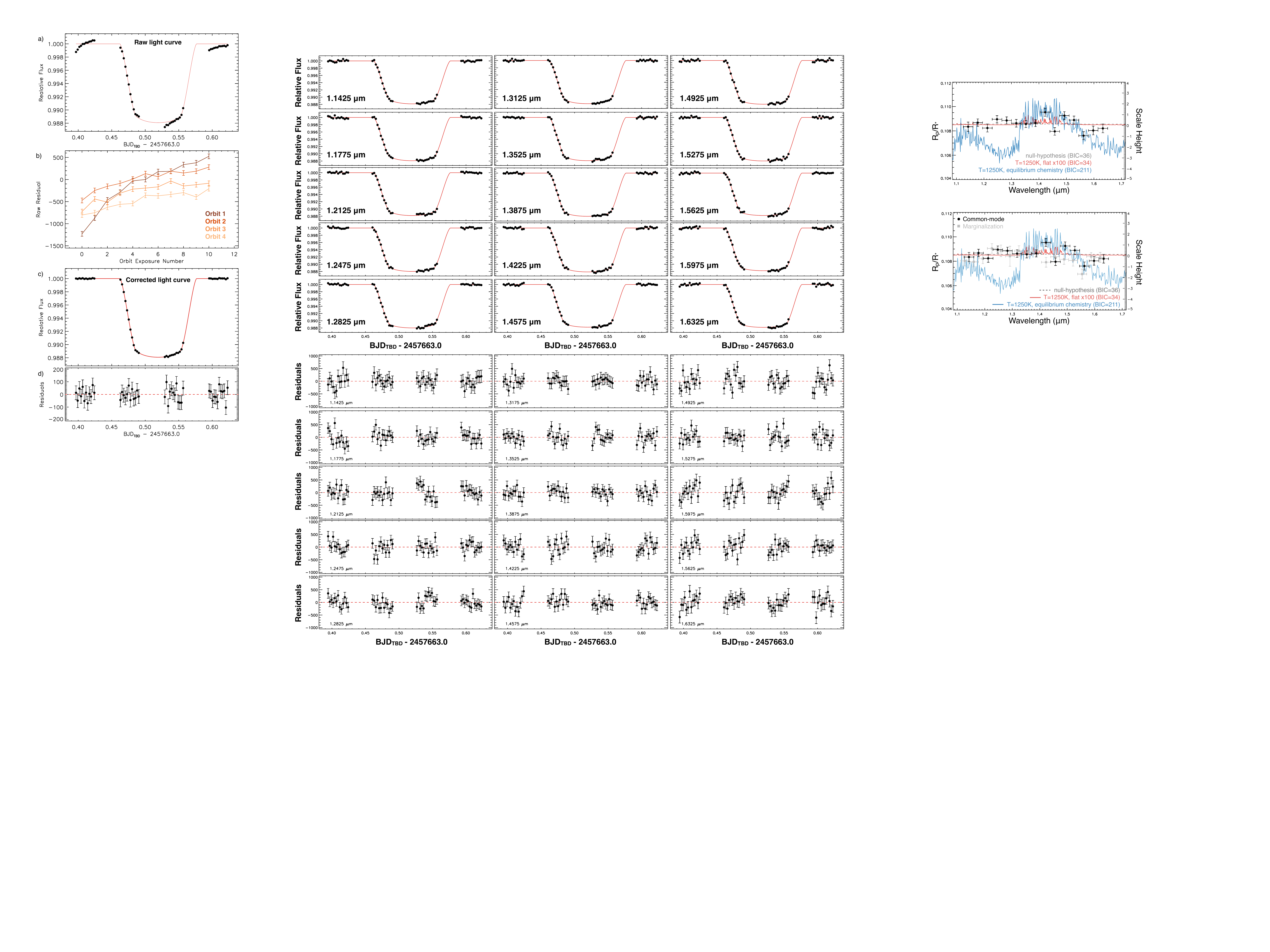}
\caption{Band-integrated light curve and residuals. a) raw light curve (black points) and model (red). b) raw orbit-to-orbit residuals showing the gradient difference between HST orbit 1 and orbits 2--4. c)light curve corrected by the systematic model and the best fit transit model (red). d) residuals in parts per million (ppm) of corrected light curve compared to the transit model (red dashed line).}
\label{fig:W101_whitelight}   
\end{figure}
The HST Panchromatic Comparative Exoplanet Treasury (PanCET) program has targeted twenty planets for the first large-scale simultaneous UVOIR comparative study of exoplanets. A major aim of PanCET is to produce the first, complete and statistically rigorous comparative study of clouds and hazes in exoplanet atmospheres over a wide range of parameters such as temperature, metallicity, mass, and radius. As part of the program WASP-101b was selected as it would have a large predicted transmission signal with strong molecular features. The planet is a highly inflated hot Jupiter (Rp\,=\,1.4\,R$_J$ and Mp\,=\,0.5\,M$_J$) in orbit around a relatively bright V Mag\,=\,10.3 F6 host star (\citealt{hellier2014}).

WASP-101b was also presented as one of the most favorable targets for JWST Early Release Science (ERS) program (\citealt{stevenson2016jwst}). The JWST ERS program focuses on testing specific observing modes in an effort to quickly supply the community with data and experience in reduction and analysis of JWST observations. As such, the targets suitable for such observations were limited to targets that fit the following criteria (\citealt{stevenson2016jwst}): Be in or near the continuous viewing zone (CVZ) of JWST (i.e. near the ecliptic poles) and have a short orbital period such that multiple transit opportunities arise and long visibility windows are open for scheduling. Additionally, the planetary system must have well constrained parameters, a relatively bright host star, and a transmission spectrum with measurable spectroscopic features. The ERS proposal submission deadline is expected to be in the summer of 2017; as such, it is vital that reconnaissance observations be carried out on the best targets in a timely manner. 

WASP-101b is expected to have a predominantly cloudy atmosphere, using the surface gravity and temperature as predictive parameters outlined in \citet{stevenson2016}. However, other observed hot Jupiters in the same parameter regime have already shown contrasting properties (\citealt{Sing2016}): WASP-17b and HD\,209458b are clear while WASP-31b is cloudy. As such, it is not yet clear to what extent simplified predictive strategies will be accurate and exceptions to general trends can bring about additional insight. 

WASP-101 was one of the first observations undertaken in the new HST cycle. In this letter we present the near-infrared observations conducted using HST and the resultant transmission section with discussion on clouds and JWST.

\begin{table}
\centering
\caption[\quad HST WFC3 transmission spectrum]{Transmission spectrum of WASP-101b measured with HST WFC3 G141 grism and analysed with using common-mode correction.}
\begin{tabular}{cccc}
\hline
\hline
Wavelength, $\lambda$ & $\Delta\lambda$ & R$_p$/R$_*$ & Uncertainty \\
($\mu$m) & ($\mu$m) & ~ & ~ \\ 
\hline
    1.1425 & 0.035 & 1.1738 & 0.0082 \\ 
    1.1775 & 0.035 & 1.1810 & 0.0067 \\
    1.2125 & 0.035 & 1.1715 & 0.0070 \\
    1.2475 & 0.035 & 1.1871 & 0.0065 \\
    1.2825 & 0.035 & 1.1850 & 0.0063 \\
    1.3175 & 0.035 & 1.1800 & 0.0068 \\
    1.3525 & 0.035 & 1.1786 & 0.0063 \\
    1.3875 & 0.035 & 1.1804 & 0.0063 \\
    1.4225 & 0.035 & 1.1997 & 0.0072 \\
    1.4575 & 0.035 & 1.1655 & 0.0067 \\
    1.4925 & 0.035 & 1.1936 & 0.0072 \\
    1.5275 & 0.035 & 1.1857 & 0.0071 \\
    1.5625 & 0.035 & 1.1579 & 0.0071 \\
    1.5975 & 0.035 & 1.1671 & 0.0077 \\
    1.6325 & 0.035 & 1.1708 & 0.0077 \\
\hline
\end{tabular}
\label{table:observation_parameters}
\vspace{10pt}
\end{table}
\begin{figure*}
\centering 
    \includegraphics[width=0.95\textwidth]{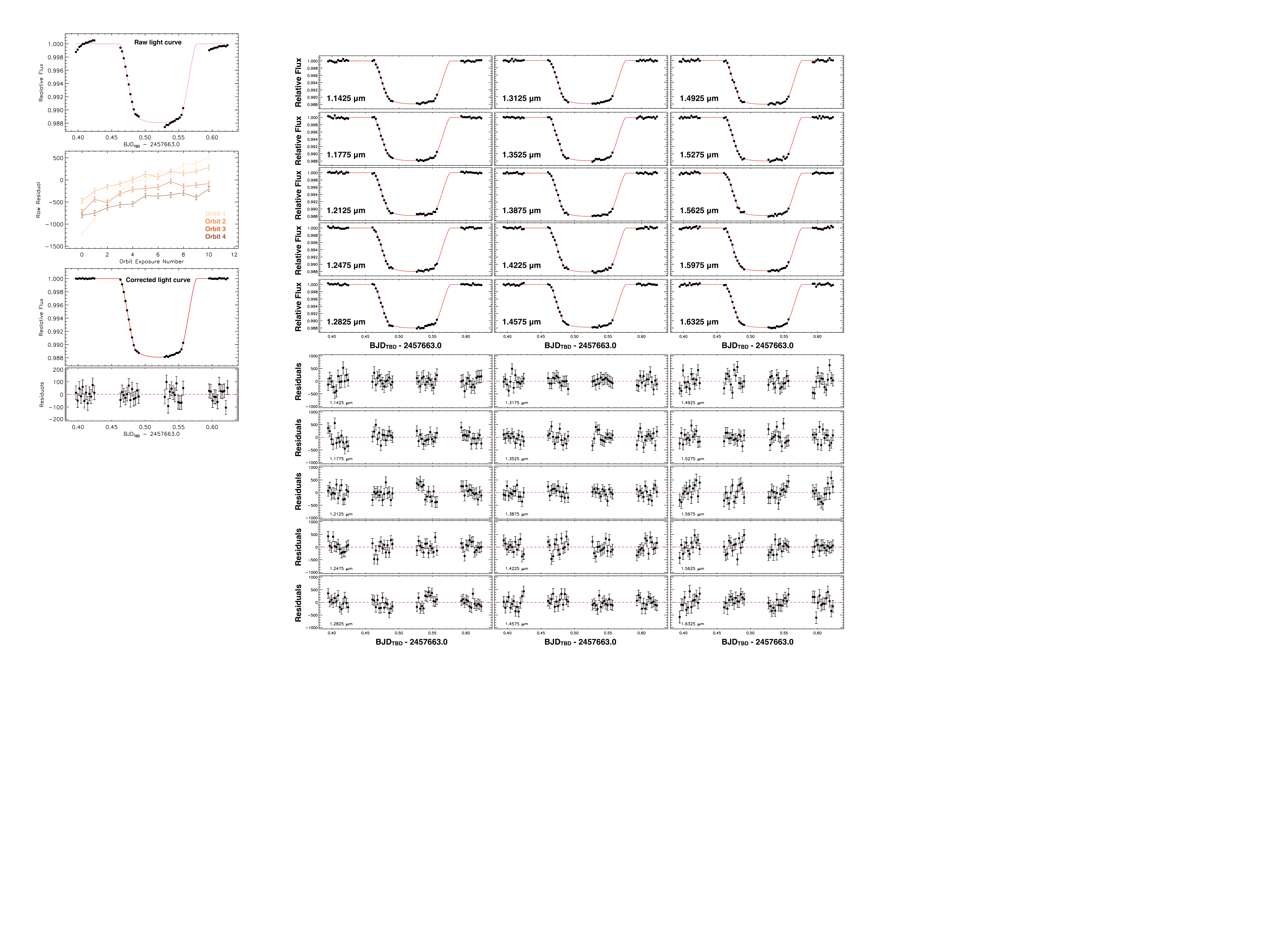}
\caption{Spectroscopic light curves and residuals for the common-mode method. Top: normalized and systematics-corrected data (points) and best-fit transit model (line) in 15 spectroscopic channels between 1.125--1.65\,$\mu$m. Bottom: corresponding residuals (ppm) from the systematic corrected light curves with their 1$\sigma$ error bars.}
\label{fig:W101_spec_lightcurves}
\end{figure*}
\section{Observations and analysis}
Observations of WASP-101 were conducted with HST WFC3 G141 grism as part of the HST PanCET program GO-14767 (PIs Sing and L\'opez-Morales) on October 2nd 2016. Observations were conducted in forward spatial scan mode, using the 512$\times$512 subarray, in SPARS25, with 7 reads per exposure, resulting in an exposure time of 138.38 seconds. We use a scan rate of $\sim$0.65 pixels per second with a final spatial scan covering $\sim$90 pixels in the cross-dispersion direction on the detector. This results in a maximum count rate of 22,000\,e$^{-}$ per pixel, which is quite low but importantly does not enter the non-linearity regime of the detector.

We use the \textit{IMA} output files from the CalWF3 pipeline which are in e$^{-}$, each are calibrated for zero-read bias, and dark current subtraction. To extract the spectra we followed a similar method to \citet{evans2016}. In each exposure the difference between sucessive non-destructive reads was taken and a top-hat filter applied around the target spectrum. The full image is then reconstructed by adding the individual reads back together. We then extract the stellar spectrum from each reconstructed image for a series of apertures. We determine the best aperture across the whole spectrum by reducing the scatter in the out-of-transit band-integrated residuals. We find a best aperture of $\pm$44 pixels for G141 around a centering profile which is found to be consistent across the spectrum for each exposure. 

\begin{figure*}
\centering 
    \includegraphics[width=0.95\textwidth]{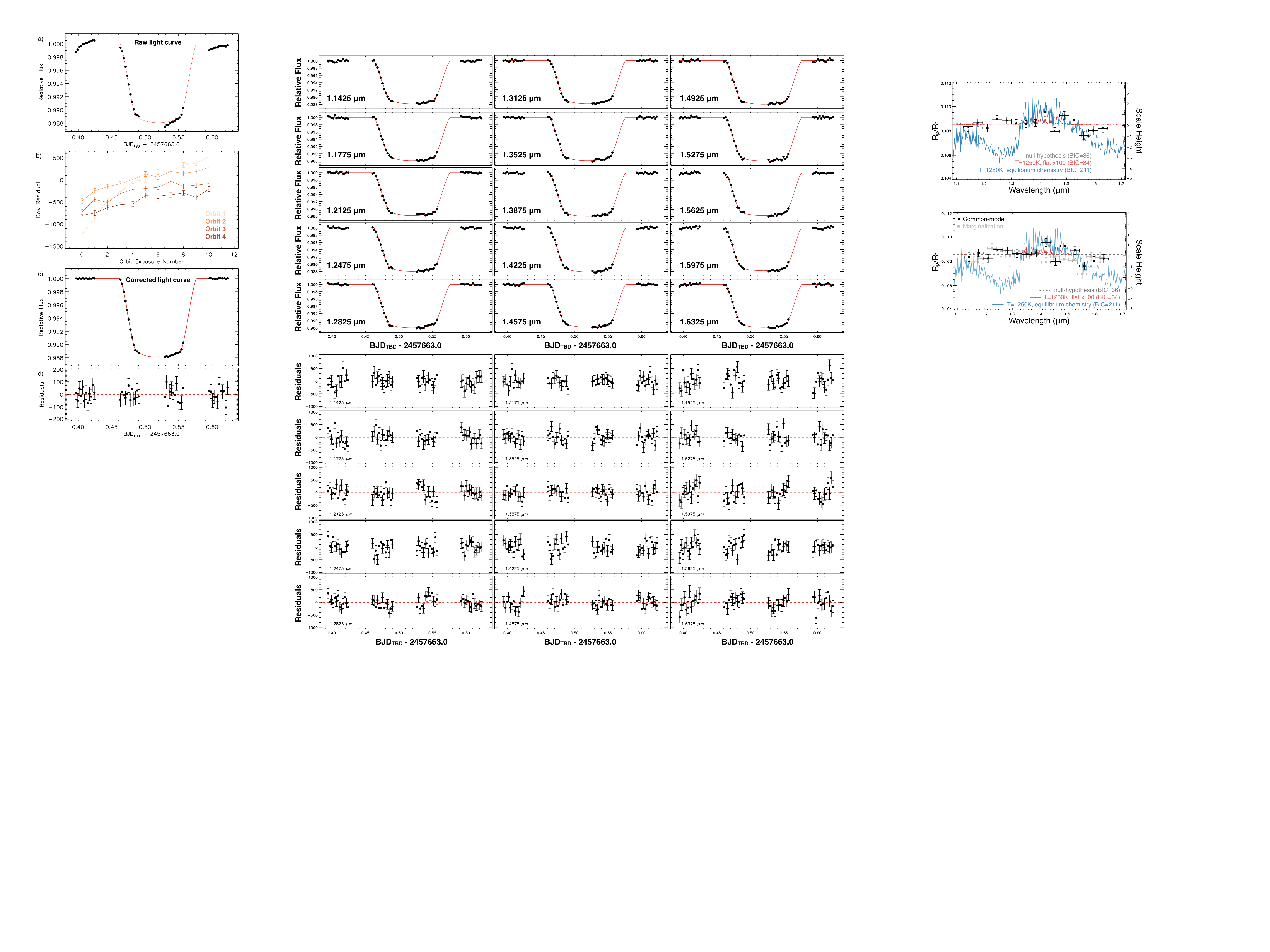}
\caption{Transmission spectrum of WASP-101b from our common-mode analysis (black points) and using marginalization (grey squares). We use the common-mode transmission spectrum and fit for the best 1D isothermal model (\citealt{fortney2010}), which has T\,=\,1250\,K and a flat cloud deck x100 the hydrogen cross-section (red). The model in blue shows the expected absorption feature for a clear atmosphere at the same temperature. The dashed line shows the null hypothesis.}
\label{fig:W101_transmission_models} 
\end{figure*}
\vspace{20pt}
\subsection{Band-integrated light curve}
We monitored the transit with HST over a total of five orbits with observations occurring prior-, during-, and post-transit. We discard the zeroth orbit as it contains vastly different systematics to the subsequent orbits. This is an effect observed in many previous studies of WFC3 exoplanet transits data (e.g., \citealt{deming2013,Mandell2013,wakeford2013,sing2013,ehrenreich2014,Sing2016}). We additionally remove the first exposure of each orbit following the buffer dump as they contain significantly lower count levels than the following exposures. This results in a total of 48 exposures over four orbits. We first analyze the band-integrated light curve, to obtain the broad-band planet-to-star radius ratio (R$_{p}$/R$_{*}$), by summing the flux between 1.125--1.65\,$\mu$m in each exposure to create the band-integrated light curve (see Fig.\ref{fig:W101_whitelight}a). 

\begin{figure}
\centering
\includegraphics[width=0.45\textwidth]{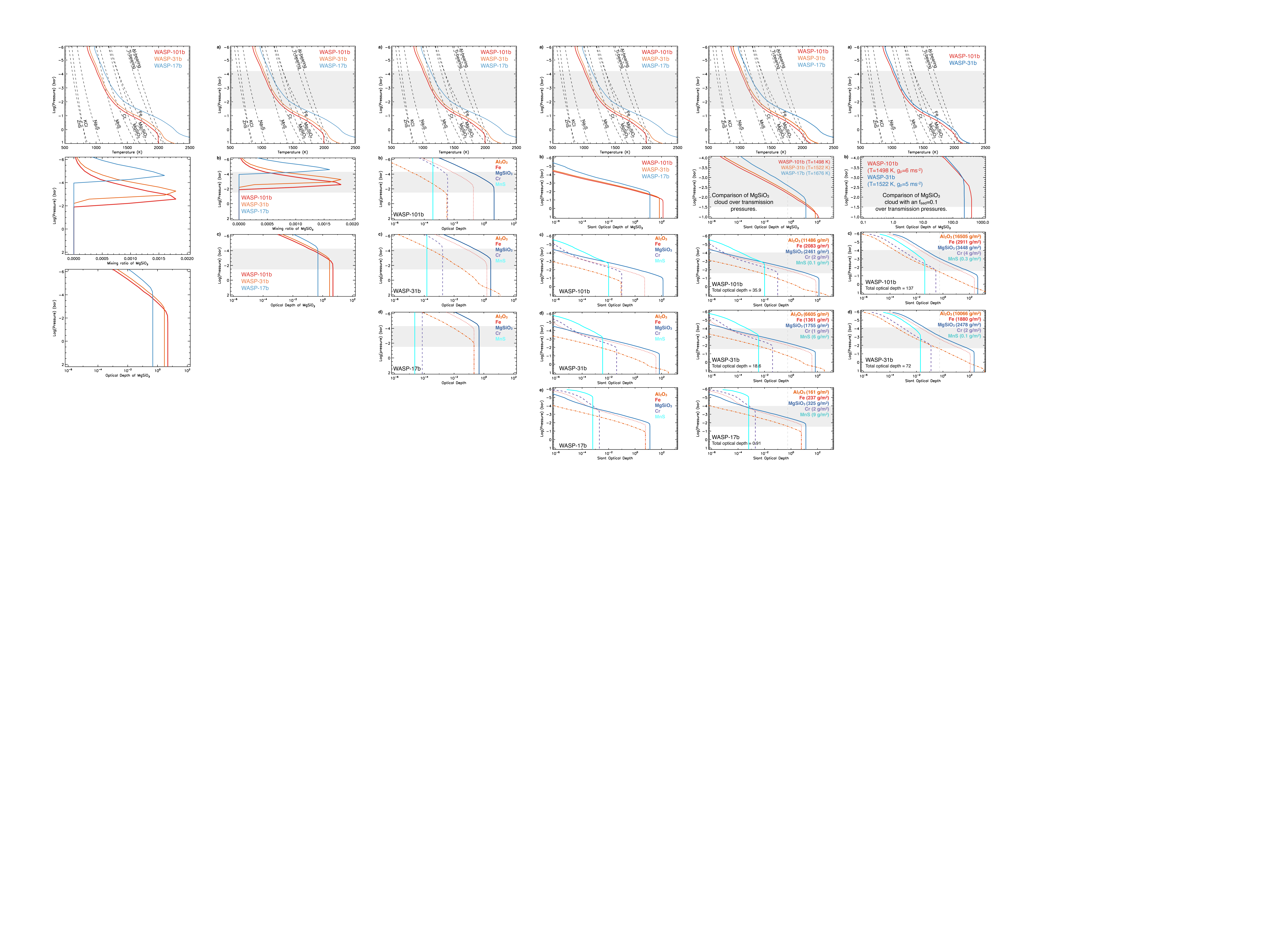}
\caption{a) Temperature-Pressure profile of WASP-101b (red, this work), with WASP-31b (blue, \citet{Sing2016}). The condensation curves (dashed labeled) indicate the different cloud species in the atmosphere where they cross the TP profile. The grey regions in each panel show the general pressure levels probed by transmission spectral observations. b) comparison between the slant optical depth of MgSiO$_3$ at the pressures probed by transmission spectroscopy. c and d) slant optical depth of each cloud forming species for a grey scattering cloud with f$_{sed}$=0.1 in each atmosphere. We use the formulations in \citet{ackerman2001} to calculate the optical depth and total optical depth of all condensates in the atmosphere with the indicated cloud mass for each species.} 
\label{fig:w101_TP}
\end{figure}
To analyze the band-integrated light curve we first have to understand the instrumental systematics impacting the observed flux. We test a number of systematic model corrections for visit-long and orbit-long trends in the data. To correct for visit-long trends we test linear, quadratic, and a rising exponential plus linear function (\citealt{stevenson2014b,line2016b}). We find that a rising exponential plus linear function in planetary phase combined with a rising exponential plus linear function in HST orbital phase produced the best fit for the data. As can be seen in Fig. \ref{fig:W101_whitelight}b the first orbit, following the removal of the zeroth orbit, exhibits a different systematic trend compared to the subsequent orbits. We see this in other unpublished datasets and appears to be related to low counts per pixel resulting from the chosen scan rates. This potentially introduces not only an upper count limit needed for HST exoplanet studies (e.g., \citealt{berta2012,wilkins2014,wakeford2016a}), but also a lower count limit that needs to be exceeded per pixel on the detector (i.e. 25,000\,e$^{-}$\,\textless\,per pixel count\,\textless\,35,000\,e$^{-}$). 

From the band-integrated light curve we measure a center of transit time of 2457663.51944\,$\pm$\,0.000080 (BJD$_{TBD}$), converted from MJD using \citet{eastman2012b}, and compute a band-integrated R$_{p}$/R$_{*}$ of 0.10860\,$\pm$\,0.00019. As the HST orbital phasing also includes the full ingress we also fit for the system inclination and the ratio of semi-major axis to stellar radius (a/R$_*$). We find an inclination of 84.73$\pm$0.05 and an improved a/R$_*$ of 8.173$\pm$0.046. We fix each parameter, excluding R$_{p}$/R$_{*}$, for each spectroscopic light curve.  

\subsection{Transmission Spectrum}
To measure the atmospheric transmission spectrum we divide the spectrum into 15 bins, each with $\Delta\lambda$\,=\,0.035\,$\mu$m, and measure R$_{p}$/R$_{*}$ in each spectroscopic light curve (Fig.\ref{fig:W101_spec_lightcurves}). We perform common-mode correction on each spectroscopic light curve using the residuals from the band-integrated light curve in addition to a linear trend in time (e.g., \citealt{deming2013,stevenson2014a}). This removes ``common'' systematics and assumes wavelength independent trends. The measured R$_{p}$/R$_{*}$ from each spectroscopic light curve are listed in Table \ref{table:observation_parameters}. We additionally measure the transmission spectrum using marginalization, which accounts for systematic trends by calculating a weight based on the evidence of fit to all tested models (see Table 2 \citealt{wakeford2016a}) and marginalizes over all results to get the measured parameter and uncertainty (\citealt{wakeford2016a}). We find both transmission spectra are in good agreement with each other (see Fig.\ref{fig:W101_transmission_models}) suggesting that there are limited or no wavelength-dependent systematics. 

We test the measured transmission spectrum against a suite of isothermal models calculated from a grid formulated in \citet{fortney2010}. Each one-dimensional isothermal model assumes solar metallicity, and includes a self-consistent treatment of radiative transfer and thermo-chemical equilibrium of neutral and ionic species. These models account for condensation and thermal ionization, though do not include photochemistry (\citealt{lodders1999}; \citealt{loddersfegley2002,lodders2002,visscher2006,loddersfegley2006,freedman2008,lodders2010}).

We find that the transmission spectrum of WASP-101b shows no evidence for H$_2$O absorption. We can rule out spectral features at 13$\sigma$ significance. As it is well known, there is degeneracy in the interpretation of a flat transmission spectrum at the WFC3 G141 wavelengths. Such flat spectra could be produced either by a cloudy atmosphere, or by an atmosphere with very low H2O content. However, based on the  recent large HST hot Jupiter survey results by \citet{Sing2016}, clouds appear to be the most likely scenario for a planet like WASP-101b. To model the lack of molecular features in the transmission spectrum we use a uniform, wavelength-independent, grey-scattering cloud with scattering approximated by increasing the Rayleigh cross-section of hydrogen at 350\,nm by 100x, which likely corresponds to clouds formed with large particle sizes (\textgreater1$\mu$m). This significantly reduces the amplitude of the expected H$_2$O absorption features. However, this has only a marginally better fit compared to the null-hypothesis of a completely featureless spectrum and is consistent with a flat spectrum at the resolution of the data (see Fig.\ref{fig:W101_transmission_models}). 

\section{Discussion}
Obscuring condensate clouds in planetary atmospheres can remove low abundance absorbers from the gas phase, obscure absorption features from gases at deeper levels (\citealt{Sing2016}), scatter incoming radiation, and add their own absorption features (\citealt{wakeford2015,wakeford2016b}). As was shown in \citet{Sing2016} approximately 50\% of measured exoplanet transmission spectra show some source of aerosol or cloud opacity in the atmosphere. 

Out of the previously well studied hot Jupiters (e.g., \citealt{deming2013,kreidberg2014b,Sing2016}) WASP-101b is most similar to WASP-31b. Both planets orbit an F6 star with an orbital period of approximately 3.5 days. WASP-101b is marginally denser with R$_p$=1.41\,R$_J$ and M$_p$=0.50\,M$_J$ and g$_p$=6.22\,ms$^{-2}$, compared to WASP-31b with R$_p$=1.53\,R$_J$ and M$_p$=0.48\,M$_J$ with a lower g$_p$=4.95\,ms$^{-2}$. Both planets also have T$_{eq}\approx$1560\,K and modest atmospheric scale heights, around 900\,km and 1100\,km respectively, making them good targets for transmission spectral characterization.

To further explore the role of cloud opacity in the atmosphere we compute a temperature-pressure (TP) profile for WASP-101b following the methods of \citet{fortney2008} and compare it to WASP-31b (Fig. \ref{fig:w101_TP}). WASP-31b was found to have a grey cloud that sits reasonably low in the atmosphere with a spectral signature that extends from the optical to the near-IR (see \citealt{Sing2015a}, and \citealt{barstow2016}). We approximate the condensate cloud forming species from multiple condensation curves from \citet{visscher2010}, \citet{morley2012}, and \citet{wakeford2016b}. The Global Circulation Model (GCM) for WASP-31b (\citealt{kataria2016}) shows that there should be little difference between eastern and western limb temperatures. We therefore use a single global TP profile for both planets as an approximation. The TP profiles of both WASP-101b (this work) and WASP-31b (\citealt{Sing2016}) cross multiple condensation curves with aluminium- and titanium-oxides deep in the atmosphere, followed by silicates, metals, and MnS in or near the observable atmosphere. 

We use the atmospheric TP profiles with the \citet{ackerman2001} cloud code, which calculates the cumulative geometric scattering optical depth by cloud particles through the atmosphere following Mie theory (e.g., \citealt{morley2012,wakeford2015,morley2015}). We convert this to the slant optical depth (\citealt{fortney2005}) for each species and compare their cloud mass. We use a sedimentation efficiency f$_{sed}$\,=\,0.1 (\citealt{morley2015}) for both atmospheres and compare the optical depth of enstatite clouds over the critical pressures probed in transmission (Fig.\ref{fig:w101_TP}b). This shows that at pressures higher than 0.001\,bar, WASP-101b has a higher optical depth, while at lower pressures higher in the atmosphere both WASP-101b and WASP-31b have similarly significant opacity caused by enstatite clouds, where small changes are likely due to the slight difference in temperature and gravity. We show the deconstruction of the total opacity of both atmospheres in Fig.\ref{fig:w101_TP}c for WASP-101b and Fig.\ref{fig:w101_TP}d for WASP-31b, with each contribution from the different cloud forming species. The cloud model shows that in both atmospheres enstatite and iron cause the highest opacity at pressures probed by transmission spectral observations. Similar to the cloud masses shown in \citet{wakeford2016b} MnS and Cr are not significant contributors to the overall opacity of the atmosphere which is dominated by Al$_2$O$_{3}$, Fe, and MgSiO$_{3}$. It will be interesting to determine if exoplanets in the same parameter regime as WASP-101b and WASP-31b will also exhibit cloudy transmission spectral features. These models can be used to approximate the cloud species and opacities expected in transmission spectral studies in preparation for JWST, however, we note that no thermal feedback from the clouds themselves are included. 

To fully characterize the cloud in the atmosphere of WASP-101b we require the information from future optical observations with HST Space Telescope Imaging Spectrograph (STIS) which will be taken as part of the HST PanCET program. These observations can be used to constrain the nature and particle size of the opacity source (e.g., \citealt{wakeford2015,Sing2016,parmentier2016,wakeford2016b}). Additionally, if the particle size is small, \textless\,1\,$\mu$m, vibrational mode features associated with the cloud composition may be distinguishable in JWST MIRI wavelengths (\citealt{wakeford2015,wakeford2016b}).

\section{Conclusion}
We present the first results from the HST PanCET program GO-14767 (PIs Sing and L\'opez-Morales). Observations of the hot Jupiter WASP-101b were conducted in the near-infrared using HST WFC3 G141 over one transit event. As one of the community targets proposed for the JWST ERS program WASP-101 was a top priority for the exoplanet community to conduct a preliminary atmospheric characterization. We measure the transmission spectrum and show that we can rule out spectral features at 13$\sigma$ significance and the planet is most likely cloudy. Therefore, with the currents observations WASP-101b does not appear to be an ideal target to be used for JWST ERS programs aimed at observing large-amplitude molecular transmission features, as it is likely to have some cloud or aerosol opacity at longer wavelengths. 

To further characterize the cloud opacity sources in the atmosphere we compare WASP-101b to the well studied hot Jupiter WASP-31b which is near identical in planetary and system parameters. We predict these twin planets both have clouds most likely composed of enstatite as it shows the highest opacity and mass. However, observations in the optical are needed to characterize the particle sizes comprising the cloud and predict any specific absorption signatures from the clouds themselves. If the clouds are in fact composed of small particles (\textless 1\,$\mu$m) then it is likely the cloud opacity will significantly reduce with increasing wavelength making molecular detection with JWST possible. 

%
\section{Acknowledgements}
This research has made use of NASAs Astrophysics Data System, and components of the IDL astronomy library.Based on observations made with the NASA/ESA Hubble Space Telescope, obtained from the Data Archive at the Space Telescope Science Institute, which is operated by the Association of Universities for Research in Astronomy, Inc., under NASA contract NAS 5-26555. These observations are associated with program GO-14767. 
H.R. Wakeford acknowledges support by an appointment to the NASA Postdoctoral Program at Goddard Space Flight Center, administered by USRA through a contract with NASA. 
D.K. Sing, N. Nikolov, and T. Evans acknowledge funding from the European Research Council under the European Union’s Seventh Framework Programme (FP7/2007-2013) / ERC grant agreement no. 336792. 
JKB is supported by a Royal Astronomical Society Research Fellowship. 
D.E. acknowledges the financial support of the National Centre for Competence in Research `PlanetS’ supported by the Swiss National Science Foundation (SNSF). 
L. Ben-Jaffel, P. Lavvas \& A. Lecavelier acknowledge support from CNES (France) under project PACES. 
The authors would like to thank the referee for their careful examination of this letter. 

{\it Facilities:} \facility{HST (WFC3)}.
%

\begin{thebibliography}{0}%
\makeatletter
\providecommand \@ifxundefined [1]{%
 \@ifx{#1\undefined}
}%
\providecommand \@ifnum [1]{%
 \ifnum #1\expandafter \@firstoftwo
 \else \expandafter \@secondoftwo
 \fi
}%
\providecommand \@ifx [1]{%
 \ifx #1\expandafter \@firstoftwo
 \else \expandafter \@secondoftwo
 \fi
}%
\providecommand \natexlab [1]{#1}%
\providecommand \enquote  [1]{``#1''}%
\providecommand \bibnamefont  [1]{#1}%
\providecommand \bibfnamefont [1]{#1}%
\providecommand \citenamefont [1]{#1}%
\providecommand \href@noop [0]{\@secondoftwo}%
\providecommand \href [0]{\begingroup \@sanitize@url \@href}%
\providecommand \@href[1]{\@@startlink{#1}\@@href}%
\providecommand \@@href[1]{\endgroup#1\@@endlink}%
\providecommand \@sanitize@url [0]{\catcode `\\12\catcode `\$12\catcode
  `\&12\catcode `\#12\catcode `\^12\catcode `\_12\catcode `\%12\relax}%
\providecommand \@@startlink[1]{}%
\providecommand \@@endlink[0]{}%
\providecommand \url  [0]{\begingroup\@sanitize@url \@url }%
\providecommand \@url [1]{\endgroup\@href {#1}{\urlprefix }}%
\providecommand \urlprefix  [0]{URL }%
\providecommand \Eprint [0]{\href }%
\providecommand \doibase [0]{http://dx.doi.org/}%
\providecommand \selectlanguage [0]{\@gobble}%
\providecommand \bibinfo  [0]{\@secondoftwo}%
\providecommand \bibfield  [0]{\@secondoftwo}%
\providecommand \translation [1]{[#1]}%
\providecommand \BibitemOpen [0]{}%
\providecommand \bibitemStop [0]{}%
\providecommand \bibitemNoStop [0]{.\EOS\space}%
\providecommand \EOS [0]{\spacefactor3000\relax}%
\providecommand \BibitemShut  [1]{\csname bibitem#1\endcsname}%
\let\auto@bib@innerbib\@empty
\end{thebibliography}%


\begin{thebibliography}{}

\bibitem[\protect\citeauthoryear{Ackerman \& Marley}{Ackerman \&
  Marley}{2001}]{ackerman2001}
Ackerman A.~S.,  Marley M.~S.,  2001, \apj, 556, 872

\bibitem[\protect\citeauthoryear{{Barstow}, {Aigrain}, {Irwin} \&
  {Sing}}{{Barstow} et~al.}{2016}]{barstow2016}
{Barstow} J.~K.,  {Aigrain} S.,  {Irwin} P.~G.~J.,    {Sing} D.~K.,  2016,
  ArXiv e-prints

\bibitem[\protect\citeauthoryear{{Benneke}}{{Benneke}}{2015}]{benneke2015}
{Benneke} B.,  2015, ArXiv e-prints

\bibitem[\protect\citeauthoryear{Berta, Charbonneau, D{\'e}sert, Kempton,
  McCullough, Burke, Fortney, Irwin, Nutzman \& Homeier}{Berta
  et~al.}{2012}]{berta2012}
Berta Z.~K.,  Charbonneau D.,  D{\'e}sert J.-M.,  Kempton E. M.-R.,  McCullough
  P.~R.,  Burke C.~J.,  Fortney J.~J.,  Irwin J.,  Nutzman P.,    Homeier D.,
  2012, \apj, 747, 35

\bibitem[\protect\citeauthoryear{Deming, Wilkins, McCullough, Burrows, Fortney,
  Agol, Dobbs-Dixon, Madhusudhan, Crouzet, Desert et~al.,}{Deming
  et~al.}{2013}]{deming2013}
Deming D.,  Wilkins A.,  McCullough P.,  Burrows A.,  Fortney J.~J.,  Agol E.,
  Dobbs-Dixon I.,  Madhusudhan N.,  Crouzet N.,  Desert J.-M.,    et~al., 2013,
  \apj, 774, 95

\bibitem[\protect\citeauthoryear{{Eastman}}{{Eastman}}{2012}]{eastman2012b}
{Eastman} J.,  2012, Astrophysics Source Code Library, p.~6012

\bibitem[\protect\citeauthoryear{{Ehrenreich}, {Bonfils}, {Lovis}, {Delfosse},
  {Forveille}, {Mayor}, {Neves}, {Santos}, {Udry} \&
  {S{\'e}gransan}}{{Ehrenreich} et~al.}{2014}]{ehrenreich2014}
{Ehrenreich} D.,  {Bonfils} X.,  {Lovis} C.,  {Delfosse} X.,  {Forveille} T.,
  {Mayor} M.,  {Neves} V.,  {Santos} N.~C.,  {Udry} S.,    {S{\'e}gransan} D.,
  2014, \aap, 570, A89

\bibitem[\protect\citeauthoryear{{Evans}, {Sing}, {Wakeford}, {Nikolov},
  {Ballester}, {Drummond}, {Kataria}, {Gibson}, {Amundsen} \& {Spake}}{{Evans}
  et~al.}{2016}]{evans2016}
{Evans} T.~M.,  {Sing} D.~K.,  {Wakeford} H.~R.,  {Nikolov} N.,  {Ballester}
  G.~E.,  {Drummond} B.,  {Kataria} T.,  {Gibson} N.~P.,  {Amundsen} D.~S.,
  {Spake} J.,  2016, \apjl, 822, L4

\bibitem[\protect\citeauthoryear{{Fortney}}{{Fortney}}{2005}]{fortney2005}
{Fortney} J.~J.,  2005, \mnras, 364, 649

\bibitem[\protect\citeauthoryear{{Fortney}, {Lodders}, {Marley} \&
  {Freedman}}{{Fortney} et~al.}{2008}]{fortney2008}
{Fortney} J.~J.,  {Lodders} K.,  {Marley} M.~S.,    {Freedman} R.~S.,  2008,
  \apj, 678, 1419

\bibitem[\protect\citeauthoryear{{Fortney}, {Mordasini}, {Nettelmann},
  {Kempton}, {Greene} \& {Zahnle}}{{Fortney} et~al.}{2013}]{fortney2013}
{Fortney} J.~J.,  {Mordasini} C.,  {Nettelmann} N.,  {Kempton} E.~M.-R.,
  {Greene} T.~P.,    {Zahnle} K.,  2013, \apj, 775, 80

\bibitem[\protect\citeauthoryear{Fortney, Shabram, Showman, Lian, Freedman,
  Marley \& Lewis}{Fortney et~al.}{2010}]{fortney2010}
Fortney J.~J.,  Shabram M.,  Showman A.~P.,  Lian Y.,  Freedman R.~S.,  Marley
  M.~S.,    Lewis N.~K.,  2010, \apj, 709, 1396

\bibitem[\protect\citeauthoryear{{Freedman}, {Marley} \& {Lodders}}{{Freedman}
  et~al.}{2008}]{freedman2008}
{Freedman} R.~S.,  {Marley} M.~S.,    {Lodders} K.,  2008, \apjs, 174, 504

\bibitem[\protect\citeauthoryear{{Greene}, {Line}, {Montero}, {Fortney},
  {Lustig-Yaeger} \& {Luther}}{{Greene} et~al.}{2016}]{greene2016}
{Greene} T.~P.,  {Line} M.~R.,  {Montero} C.,  {Fortney} J.~J.,
  {Lustig-Yaeger} J.,    {Luther} K.,  2016, \apj, 817, 17

\bibitem[\protect\citeauthoryear{{Hellier}, {Anderson}, {Cameron}, {Delrez},
  {Gillon}, {Jehin}, {Lendl}, {Maxted}, {Pepe}, {Pollacco}, {Queloz},
  {S{\'e}gransan}, {Smalley}, {Smith}, {Southworth}, {Triaud}, {Udry} \&
  {West}}{{Hellier} et~al.}{2014}]{hellier2014}
{Hellier} C.,  {Anderson} D.~R.,  {Cameron} A.~C.,  {Delrez} L.,  {Gillon} M.,
  {Jehin} E.,  {Lendl} M.,  {Maxted} P.~F.~L.,  {Pepe} F.,  {Pollacco} D.,
  {Queloz} D.,  {S{\'e}gransan} D.,  {Smalley} B.,  {Smith} A.~M.~S.,
  {Southworth} J.,  {Triaud} A.~H.~M.~J.,  {Udry} S.,    {West} R.~G.,  2014,
  \mnras, 440, 1982

\bibitem[\protect\citeauthoryear{{Kataria}, {Sing}, {Lewis}, {Visscher},
  {Showman}, {Fortney} \& {Marley}}{{Kataria} et~al.}{2016}]{kataria2016}
{Kataria} T.,  {Sing} D.~K.,  {Lewis} N.~K.,  {Visscher} C.,  {Showman} A.~P.,
  {Fortney} J.~J.,    {Marley} M.~S.,  2016, \apj, 821, 9

\bibitem[\protect\citeauthoryear{{Knutson}, {Benneke}, {Deming} \&
  {Homeier}}{{Knutson} et~al.}{2014}]{knutson2014a}
{Knutson} H.~A.,  {Benneke} B.,  {Deming} D.,    {Homeier} D.,  2014, \nat,
  505, 66

\bibitem[\protect\citeauthoryear{{Kreidberg}, {Bean}, {D{\'e}sert}, {Line},
  {Fortney}, {Madhusudhan}, {Stevenson}, {Showman}, {Charbonneau},
  {McCullough}, {Seager}, {Burrows}, {Henry}, {Williamson}, {Kataria} \&
  {Homeier}}{{Kreidberg} et~al.}{2014}]{kreidberg2014b}
{Kreidberg} L.,  {Bean} J.~L.,  {D{\'e}sert} J.-M.,  {Line} M.~R.,  {Fortney}
  J.~J.,  {Madhusudhan} N.,  {Stevenson} K.~B.,  {Showman} A.~P.,
  {Charbonneau} D.,  {McCullough} P.~R.,  {Seager} S.,  {Burrows} A.,  {Henry}
  G.~W.,  {Williamson} M.,  {Kataria} T.,    {Homeier} D.,  2014, \apjl, 793,
  L27

\bibitem[\protect\citeauthoryear{{Line}, {Knutson}, {Deming}, {Wilkins} \&
  {Desert}}{{Line} et~al.}{2013}]{line2013}
{Line} M.~R.,  {Knutson} H.,  {Deming} D.,  {Wilkins} A.,    {Desert} J.-M.,
  2013, \apj, 778, 183

\bibitem[\protect\citeauthoryear{{Line}, {Stevenson}, {Bean}, {Desert},
  {Fortney}, {Kreidberg}, {Madhusudhan}, {Showman} \& {Diamond-Lowe}}{{Line}
  et~al.}{2016}]{line2016b}
{Line} M.~R.,  {Stevenson} K.~B.,  {Bean} J.,  {Desert} J.-M.,  {Fortney}
  J.~J.,  {Kreidberg} L.,  {Madhusudhan} N.,  {Showman} A.~P.,
  {Diamond-Lowe} H.,  2016, ArXiv e-prints

\bibitem[\protect\citeauthoryear{Lodders}{Lodders}{1999}]{lodders1999}
Lodders K.,  1999, \apj, 519, 793

\bibitem[\protect\citeauthoryear{{Lodders}}{{Lodders}}{2002}]{lodders2002}
{Lodders} K.,  2002, \apj, 577, 974

\bibitem[\protect\citeauthoryear{{Lodders}}{{Lodders}}{2010}]{lodders2010}
{Lodders} K.,  2010, {Exoplanet Chemistry}.
p.~157

\bibitem[\protect\citeauthoryear{{Lodders} \& {Fegley}}{{Lodders} \&
  {Fegley}}{2002}]{loddersfegley2002}
{Lodders} K.,  {Fegley} B.,  2002, \icarus, 155, 393

\bibitem[\protect\citeauthoryear{{Lodders} \& {Fegley} Jr. B. edited
  by~{Mason}}{{Lodders} \& {Fegley}}{2006}]{loddersfegley2006}
{Lodders} K.,  {Fegley} Jr. B. edited by~{Mason} J.~W.,  2006, {Chemistry of
  Low Mass Substellar Objects}.
Springer Praxis, p.~1

\bibitem[\protect\citeauthoryear{{Mandell}, {Haynes}, {Sinukoff},
  {Madhusudhan}, {Burrows} \& {Deming}}{{Mandell} et~al.}{2013}]{Mandell2013}
{Mandell} A.~M.,  {Haynes} K.,  {Sinukoff} E.,  {Madhusudhan} N.,  {Burrows}
  A.,    {Deming} D.,  2013, \apj, 779, 128

\bibitem[\protect\citeauthoryear{Morley, Fortney, Marley, Visscher, Saumon \&
  Leggett}{Morley et~al.}{2012}]{morley2012}
Morley C.~V.,  Fortney J.~J.,  Marley M.~S.,  Visscher C.,  Saumon D.,
  Leggett S.,  2012, \apj, 756, 172

\bibitem[\protect\citeauthoryear{{Morley}, {Fortney}, {Marley}, {Zahnle},
  {Line}, {Kempton}, {Lewis} \& {Cahoy}}{{Morley} et~al.}{2015}]{morley2015}
{Morley} C.~V.,  {Fortney} J.~J.,  {Marley} M.~S.,  {Zahnle} K.,  {Line} M.,
  {Kempton} E.,  {Lewis} N.,    {Cahoy} K.,  2015, \apj, 815, 110

\bibitem[\protect\citeauthoryear{{Parmentier}, {Fortney}, {Showman}, {Morley}
  \& {Marley}}{{Parmentier} et~al.}{2016}]{parmentier2016}
{Parmentier} V.,  {Fortney} J.~J.,  {Showman} A.~P.,  {Morley} C.~V.,
  {Marley} M.~S.,  2016, ArXiv e-prints

\bibitem[\protect\citeauthoryear{Sing, Fortney, Nikolov, Wakeford, Kataria,
  Evans, Aigrain, Ballester, Burrows, Deming, D{\'e}sert, Gibson, Henry,
  Huitson, Knutson, Etangs, Pont, Showman, Vidal-Madjar, Williamson \&
  Wilson}{Sing et~al.}{2016}]{Sing2016}
Sing D.~K.,  Fortney J.~J.,  Nikolov N.,  Wakeford H.~R.,  Kataria T.,  Evans
  T.~M.,  Aigrain S.,  Ballester G.~E.,  Burrows A.~S.,  Deming D.,  D{\'e}sert
  J.-M.,  Gibson N.~P.,  Henry G.~W.,  Huitson C.~M.,  Knutson H.~A.,  Etangs
  A. L.~d.,  Pont F.,  Showman A.~P.,  Vidal-Madjar A.,  Williamson M.~H.,
  Wilson P.~A.,  2016, Nature, 529, 59

\bibitem[\protect\citeauthoryear{{Sing}, {Lecavelier des Etangs}, {Fortney},
  {Burrows}, {Pont}, {Wakeford}, {Ballester}, {Nikolov}, {Henry} \& {et
  al.}}{{Sing} et~al.}{2013}]{sing2013}
{Sing} D.~K.,  {Lecavelier des Etangs} A.,  {Fortney} J.~J.,  {Burrows} A.~S.,
  {Pont} F.,  {Wakeford} H.~R.,  {Ballester} G.~E.,  {Nikolov} N.,  {Henry}
  G.~W.,    {et al.} 2013, \mnras, 436, 2956

\bibitem[\protect\citeauthoryear{{Sing}, {Wakeford}, {Showman}, {Nikolov},
  {Fortney}, {Burrows}, {Ballester}, {Deming} \& {et al.}}{{Sing}
  et~al.}{2015}]{Sing2015a}
{Sing} D.~K.,  {Wakeford} H.~R.,  {Showman} A.~P.,  {Nikolov} N.,  {Fortney}
  J.~J.,  {Burrows} A.~S.,  {Ballester} G.~E.,  {Deming} D.,    {et al.} 2015,
  \mnras, 446, 2428

\bibitem[\protect\citeauthoryear{{Stevenson}}{{Stevenson}}{2016}]{stevenson2016}
{Stevenson} K.~B.,  2016, \apjl, 817, L16

\bibitem[\protect\citeauthoryear{{Stevenson}, {Bean}, {Seifahrt}, {D{\'e}sert},
  {Madhusudhan}, {Bergmann}, {Kreidberg} \& {Homeier}}{{Stevenson}
  et~al.}{2014}]{stevenson2014a}
{Stevenson} K.~B.,  {Bean} J.~L.,  {Seifahrt} A.,  {D{\'e}sert} J.-M.,
  {Madhusudhan} N.,  {Bergmann} M.,  {Kreidberg} L.,    {Homeier} D.,  2014,
  \aj, 147, 161

\bibitem[\protect\citeauthoryear{{Stevenson}, {D{\'e}sert}, {Line}, {Bean},
  {Fortney}, {Showman}, {Kataria}, {Kreidberg}, {McCullough}, {Henry},
  {Charbonneau}, {Burrows}, {Seager}, {Madhusudhan}, {Williamson} \&
  {Homeier}}{{Stevenson} et~al.}{2014}]{stevenson2014b}
{Stevenson} K.~B.,  {D{\'e}sert} J.-M.,  {Line} M.~R.,  {Bean} J.~L.,
  {Fortney} J.~J.,  {Showman} A.~P.,  {Kataria} T.,  {Kreidberg} L.,
  {McCullough} P.~R.,  {Henry} G.~W.,  {Charbonneau} D.,  {Burrows} A.,
  {Seager} S.,  {Madhusudhan} N.,  {Williamson} M.~H.,    {Homeier} D.,  2014,
  Science, 346, 838

\bibitem[\protect\citeauthoryear{{Stevenson}, {Lewis}, {Bean}, {Beichman},
  {Fraine}, {Kilpatrick}, {Krick}, {Lothringer} \& {et al.}}{{Stevenson}
  et~al.}{2016}]{stevenson2016jwst}
{Stevenson} K.~B.,  {Lewis} N.~K.,  {Bean} J.~L.,  {Beichman} C.,  {Fraine} J.,
   {Kilpatrick} B.~M.,  {Krick} J.~E.,  {Lothringer} J.~D.,    {et al.} 2016,
  \pasp, 128, 094401

\bibitem[\protect\citeauthoryear{{Visscher}, {Lodders} \& {Fegley}
  Jr.}{{Visscher} et~al.}{2006}]{visscher2006}
{Visscher} C.,  {Lodders} K.,    {Fegley} Jr. B.,  2006, \apj, 648, 1181

\bibitem[\protect\citeauthoryear{{Visscher}, {Lodders} \& {Fegley}
  Jr.}{{Visscher} et~al.}{2010}]{visscher2010}
{Visscher} C.,  {Lodders} K.,    {Fegley} Jr. B.,  2010, \apj, 716, 1060

\bibitem[\protect\citeauthoryear{Wakeford, Sing, Deming, Gibson, Fortney,
  Burrows, Ballester, Nikolov, Aigrain, Henry et~al.,}{Wakeford
  et~al.}{2013}]{wakeford2013}
Wakeford H.,  Sing D.,  Deming D.,  Gibson N.,  Fortney J.,  Burrows A.,
  Ballester G.,  Nikolov N.,  Aigrain S.,  Henry G.,    et~al., 2013, \mnras,
  435, 3481

\bibitem[\protect\citeauthoryear{{Wakeford} \& {Sing}}{{Wakeford} \&
  {Sing}}{2015}]{wakeford2015}
{Wakeford} H.~R.,  {Sing} D.~K.,  2015, \aap, 573, A122

\bibitem[\protect\citeauthoryear{Wakeford, Sing, Evans, Deming \&
  Mandell}{Wakeford et~al.}{2016}]{wakeford2016a}
Wakeford H.~R.,  Sing D.~K.,  Evans T.,  Deming D.,    Mandell A.,  2016, The
  Astrophysical Journal, 819, 10

\bibitem[\protect\citeauthoryear{{Wakeford}, {Visscher}, {Lewis}, {Kataria},
  {Marley}, {Fortney} \& {Mandell}}{{Wakeford} et~al.}{2017}]{wakeford2016b}
{Wakeford} H.~R.,  {Visscher} C.,  {Lewis} N.~K.,  {Kataria} T.,  {Marley}
  M.~S.,  {Fortney} J.~J.,    {Mandell} A.~M.,  2017, \mnras, 464, 4247

\bibitem[\protect\citeauthoryear{{Wilkins}, {Deming}, {Madhusudhan}, {Burrows},
  {Knutson}, {McCullough} \& {Ranjan}}{{Wilkins} et~al.}{2014}]{wilkins2014}
{Wilkins} A.~N.,  {Deming} D.,  {Madhusudhan} N.,  {Burrows} A.,  {Knutson} H.,
   {McCullough} P.,    {Ranjan} S.,  2014, \apj, 783, 113

\end{thebibliography}

\end{document}